\input phyzzx.tex
\tolerance=1000
\voffset=-0.0cm
\hoffset=0.7cm
\sequentialequations
\def\rl{\rightline}

\def\t1{{\tilde 1}}

\def\t{\theta}

\REF{\COS}{G. Dvali, R. Kallosh and A. Van Proeyen, hep-th/0312005.}
\REF{\POL}{E. J. Copeland, R. C. Myers and J. Polchinski, hep-th/0312067.}
\REF{\DTE}{E. Halyo, Phys. Lett. {\bf B387} (1996) 43, hep-ph/9606423.} 
\REF{\BIN}{P. Binetruy and G. Dvali, Phys. Lett. {\bf B450} (1996) 241, hep-ph/9606342.}
\REF{\TYP}{E. Halyo, Phys. Lett. {\bf B454} (1999) 223, hep-ph/9901302.}
\REF{\KLE}{I. R. Klebanov and E. Witten, Nucl. Phys. {\bf B536} (1998) 199, hep-th/9807080.}
\REF{\MUK}{K. Dasgupta and S. Mukhi, hep-th/9811139.}
\REF{\DGM}{M. R. Douglas, B. R. Greene and D. R. Morrison, Nucl. Phys. {\bf B506} (1997) 84, hep-th/9704151.}
\REF{\HKO}{C. P. Herzog, I. R. Klebanov and P. Ouyag, hep-th/0205100.}
\REF{\MAL}{J. Maldacena and C. Nunez, Phys. Rev. Lett. {\bf 88} (2001) 588, hep-th/0008001; Int. Journ. Mod. Phys. {\bf A16} (2001) 822, hep-th/0007018.}
\REF{\BER}{M. Bertolini, hep-th/0303160 and references therein.} 
\REF{\ZAF}{F. Bigazzi, A. L. Cotrone, M. Petrini and A. Zaffaroni, hep-th/0303319 and references therein.}
\REF{\DSS}{G. Dvali, Q. Shafi and S. Solganik, hep-th/0105203.}
\REF{\BUR}{C. P. Burgess at. al. JHEP {\bf 07} (2001) 047, hep-th/0105204.}
\REF{\ALE}{S. H. Alexander, Phys. Rev {\bf D65} (2002) 023507, hep-th/0105032.}
\REF{\EDI}{E. Halyo, hep-ph/0105341.}
\REF{\SHI}{G. Shiu and S.-H. H. Tye, Phys. Lett. {\bf B516} (2001) 421, hep-th/0106274.}
\REF{\CAR}{C. Herdeiro, S. Hirano and R. Kallosh, JHEP {\bf 0112} (2001) 027, hep-th/0110271.}
\REF{\SHA}{B. S. Kyae and Q. Shafi, Phys. Lett. {\bf B526} (2002) 379, hep-ph/0111101.}
\REF{\BEL}{J. Garcia-Bellido, R. Rabadan and F. Zamora, JHEP {\bf 01} (2002) 036, hep-th/0112147.}
\REF{\KAL}{K. Dasgupta, C. Herdeiro, S. Hirano and R. Kallosh, Phys. Rev. {\bf D65} (2002) 126002, hep-th/0203019.}
\REF{\JST}{N. T. Jones, H. Stoica and S. H. Tye, JHEP {\bf 0207} (2002) 051, hep-th/0203163.}
\REF{\OTH}{T. Matsua, hep-th/0302035; hep-th/0302078; T. Sato, hep-th/0304237; Y. Piao, X. Zhang and Y. Zhang, hep-th/0305171; C. P. Burgess, 
J. Cline and R. Holman, hep-th/0306079.}
\REF{\LAS}{E. Halyo, hep-th/0307223.}
\REF{\TYE}{S. Sarangi and S. H. Tye, Phys. Lett. {\bf B536} (2002) 185, hep-th/204074; N. T. Jones, H. Stoica and S. H. Tye, Phys. Lett. {\bf B563} (2003) 6, hep-th/0303269;
G. Dvali and A. Vilenkin, hep-th/0312007.}
\REF{\JAU}{D. E. Diaconescu, M. R. Douglas and J. Gomis, JHEP {\bf 9802} (1998) 013, hep-th/9712230.}
\REF{\BLO}{E. Halyo, hep-th/0312042.} 
\REF{\NEW}{E. Halyo, ``D--Brane Inflation from Resolved Conifolds'' in preparation.}
\REF{\QUI}{M. R. Douglas and G. W. Moore, hep-th/9603167; C. V. Johnson and R. C. Myers, hep-th/9610140.}

\singlespace
\rl{hep-th/0312268}
\rl{\today}
\pagenumber=0
\normalspace
\medskip
\bigskip
\titlestyle{\bf{Cosmic D--term Strings as Wrapped D3 Branes}}
\smallskip
\author{ Edi Halyo{\footnote*{e--mail address: vhalyo@stanford.edu}}}
\smallskip
 \centerline{California Institute for Physics and Astrophysics}
\centerline{366 Cambridge St.}
\centerline{Palo Alto, CA 94306}
\smallskip
\vskip 2 cm
\titlestyle{\bf ABSTRACT}

We describe cosmic D--term strings as D3 branes wrapped on a resolved conifold. The matter content that gives rise to D--term strings is shown to describe the world--volume theory of a 
space--filling D3 brane transverse to the conifold which itself is a wrapped D5 brane. We show that, in this brane theory, the tension of the wrapped D3 branes matches that of the D--term 
strings. We argue that there is a new type of cosmic string which arises from fractional D1 branes on the world--volume of a fractional D3 brane.

\singlespace
\vskip 0.5cm
\endpage
\normalspace

\centerline{\bf 1. Introduction}
\medskip

Recently, a solution that corresponds to a cosmic string, called a D--term string[\COS] was obtained in a ${\cal N}=1$ supergravity theory with a $U(1)$ gauge group and a charged complex scalar.
It was shown that D--term strings exist only when there is an anomalous D--term in the scalar potential. This string solution is very similar to the usual Nielsen--Olesen vortex with a
string tension proportional to the anomalous D--term.

In this letter, we show that D--term strings can be understood as D3 branes wrapped over the $P^1$ of a resolved conifold. In order to show this, we consider IIB string theory with a D3 brane 
(which itself is a D5 brane wrapped over the same $P^1$) at a conifold singularity of a compact 
Calabi--Yau manifold. The compact transverse space allows us to have gravity on the brane. The field theory on the brane world--volume is an ${\cal N}=1$ with a $U(1)$ gauge group and 
two complex scalars
(one of which is irrelevant for our purposes). The superpotential of this theory vanishes at the tree level. The anomalous D--term on the world--volume theory corresponds to the resolution 
of the conifold singularity, i.e.
the replacement of the singularity by a $P^1$ of finite radius. A D3 brane wrapped on this $P^1$, will be a string on the $3+1$ dimensional world--volume theory. 

We show that the tension of this string (which is proportional to the blow--up radius) matches that of D--term strings (which is proportional to the anomalous D--term). In addition, this 
string carries one unit of gauge flux just like a vortex. Therefore, 
we identify D--term strings with D3 branes wrapped on resolved conifold singularities as they are seen in the 
world--volume of D3 branes transverse to the conifold. This agrees nicely with the arguments in ref. [\POL] which state that cosmic F or D strings cannot be stable if there are $M^4$--filling 
branes as in our case. Thus, wrapped D3 branes appear as a new type of stable cosmic string namely D--term strings in this theory.
We argue that fractional D1 branes should also correspond to stable cosmic strings. They arise when the singularity is not resolved but there is a nonzero B--flux through the vanishing 
cycle which leads to a nonzero string tension.
These cannot be D--term strings since in this case there is no anomalous D--term in the field theory. Thus, they must be a new type of cosmic string which we call fractional cosmic
strings. They are not visible in the supergravity/field theory limit since their origin is a purely string theoretical.

This letter is organized as follows. In the next section we briefly describe D--term strings. In section 3, we describe the wrapped D3 branes, calculate their tension and identify them
with cosmic D--term strings. Section 4 includes a discussion of our results and our conclusions.

\bigskip
\centerline{\bf 2. D--term Strings}
\medskip

In ref. [\COS], a cosmic string solution was obtained in an ${\cal N}=1$ supergravity model in which the matter sector comprised of a $U(1)$ gauge boson and a (negatively) charged 
complex scalar with 
vanishing superpotential. In addition, an anomalous D--term, $\xi$, which is crucial for the existence of the string solution was included. The bosonic Lagrangian of this model is
$${\cal L}=-M_P^2 R - \partial_{\mu} \phi \partial ^{\mu} \phi -{1 \over 4} F_{\mu \nu} F^{\mu \nu}- g^2(-|\phi|^2+\xi)^2 \eqno(1)$$ 
The scalar potential above is the one that appears in models of D--term inflation[\DTE-\TYP] with some scalars set to zero.
It was shown that there is a string solution which (far away from its core) at large $r$ is given by
$$\phi=\sqrt \xi \qquad \qquad  W_{\theta}={n \over g} d\theta \qquad \qquad F_{\mu \nu}=0 \eqno(2)$$
We see that away from the string $D=0$ and supersymmetry is restored. As usual, the string creates a locally flat conical metric
$$ds^2=-dt^2+dz^2+dr^2+r^2 \left(1-{{n \xi} \over {M_P^2}} \right) d\theta^2 \eqno(3)$$
Near the core of the string (at small $r$) the solution is given by
$$\phi=0 \qquad \qquad W_{\theta}={M_P^2 \over {g \xi}} \left(1-cos({{g \xi} \over M_P} r) \right) \qquad \qquad F_{r \theta}=M_P sin({{g \xi} \over M_P}r) \eqno(4)$$
In this limit half the supersymmetries are broken due to the nonvanishing curvature from the metric
$$ds^2=-dt^2+dz^2+dr^2+\left({M_P \over {g \xi}} sin({{g \xi} \over M_P}r) \right)^2 d\theta^2 \eqno(5)$$
The tension of D--term strings is found to be 
$$T_{st}=2 \pi n \xi \eqno(6)$$
This is a very surprising result since the tension seems to be independent of the gauge coupling $g$. One would expect for a solitonic object $T \sim 1/g^2$. We will see below that this
puzzle is resolved by wrapped branes.

\bigskip
\centerline{\bf 3. D--term Strings as Wrapped D3 Branes}
\medskip

The vortex--like solution of ref.[\COS] was obtained in an effective low--energy supergravity/field theory. In this section we would like to identify the D--term string in IIB string theory. 
In general, in IIB string theory we can have a number of one--dimensional objects. These include the F and D strings and also Dp--branes wrapped on a $p-1$ cycle of a compact manifold. 
We would in particular like to obtain a string theoretical description of D--term strings on the world--volume of a space--filling brane. This can be a D3 brane or a D(3+q)--brane wrapped on
a $q$ dimensional internal cycle.

We will first obtain the low--energy theory described by the Lagrangian in eq. (1). Consider IIB string theory with a D3 brane at a conifold singularity of a transverse, 
compact Calabi--Yau threefold. The conifold is defined by
$$z_1^2+z_2^2+z_3^2+z_4^2=0 \eqno(7)$$
which can also be written as
$$z_1^2+z_2^2+z_3^2=-z_4^2 \eqno(8)$$
In this form, the conifold is described as a fibration where the base is the $z_4$ plane (or the compactified sphere)
and the fiber is a $Z_2$ ALE space with a size that varies linearly with $|z_4|$. Let us define $z_4=X_4+iX_5$ and take the ALE space to be along the $X_{6,7,8,9}$ directions. 
In the orbifold limit, the ALE space is defined by the $Z_2$ identification $(X_6,X_7,X_8,X_9)=-(X_6,X_7,X_8,X_9)$. Thus there is a fixed point (or fixed plane given by $z_4$) at 
$X_6=X_7=X_8=X_9=0$. In addition, the size of the whole ALE space vanishes at $X_4=X_5=0$ which gives the conifold singularity. 

Now consider a D3 brane transverse to the conifold, along the $X_{1,2,3}$ directions. Since the space transverse to the brane is compact, there is four dimensional gravity on the brane
world--volume with $M_P^2 \sim V_{CY}/g_s^2 \ell_s^8$ where $V_{CY}$ is the volume of the Calabi--Yau manifold. The conifold breaks supersymmetry to $1/4$ and the brane breaks an additional 
$1/2$; as a result we get ${\cal N}=1$ supersymmetry 
on the brane. If we have one brane, the gauge group is $U(1) \times U(1)$ whereas the matter content is given by two pairs of bifundamental chiral multiplets, $A_{1,2}$ an $B_{1,2}$ 
with charges $(1,-1)$ and $(-1,1)$ respectively[\KLE].

The superpotential of this theory vanishes at tree level[\MUK,\HKO]. In order to see this, consider the D3 brane transverse to $T^2 \times K3$. In this case, the world--volume theory has
${\cal N}=2$ supersymmetry and an additional pair of gauge singlets $\Phi, \bar \Phi$ (coming from the ${\cal N}=2$ supersymmetric gauge multiplets). This theory has the superpotential
$$W(A_i,B_i,\Phi, \bar \Phi)=g~Tr \Phi(A_1B_1+A_2+B_2)+ g~Tr \bar \Phi(B_1A_1+B_2A_2) \eqno(9)$$ 
Going to the conifold from $T^2 \times ALE$ corresponds to deforming the $T^2$ to a sphere $P^1$ on which the size of the ALE space varies. This deformation is described in field theory by 
giving a(n infinite) mass to the gauge singlets, $\Phi, \bar \Phi$. These need to decouple since they are absent in the brane theory on the conifold. 
The mass terms are given by the superpotential
$$W_m={1 \over 2}m (Tr \Phi^2+ Tr \bar \Phi^2) \eqno(10)$$
Integrating the singlets out gives
$$W(A_i,B_i)={1 \over {2m}}Tr[(A_1B_1A_2B_2)-Tr(B_1A_1B_2A_2)] \eqno(11)$$
which vanishes for $m \to \infty$ in the limit that corresponds to the conifold.
The sum of the $U(1)$'s, $(1/2)[U(1)_1+U(1)_2]$ is irrelevant for our purposes ($A_i$ and $B_i$ are neutral under it) and can be neglected. Under the orthogonal combination 
$(1/2)[U(1)_1-U(1)_2]$ the $A_i$ and $B_i$
have charges of $1$ and $-1$ respectively. We can set the $A_i$'s and one of the $B_i$ to zero for our purposes. This gives us precisely the matter content of the theory in eq. (1).

Thus we have almost the theory described by eq. (1) on the D3 brane world--volume. The only missing ingredient is the anomalous D--term. However, it is well--known that the resolution
of the orbifold singularity of the ALE fiber at $X_6=X_7=X_8=X_9=0$ (and at $X_4=X_5=0$) by blowing it up by a sphere of radius $R$, correponds to an anomalous D--term on the brane 
world--volume[\DGM]. A $Z_2$ ALE space at the orbifold limit is defined by
$$z_1^2+z_2^2+z_3^2=0 \eqno(12)$$
We can resolve the singularity by blowing up the a sphere of radius $R$. This is described by
$$z_1^2+z_2^2+z_3^2=R^2 \eqno(13)$$
If we write $z_i=x_i+iy_i$ then eq. (13) becomes
$$x_i^2-y_i^2=R^2 \qquad \qquad x_i y_i=0 \eqno(14)$$ 
Now define $r^2=x_i^2+y_i^2$. Then for $r^2=R^2$ we need to have $y_i=0$ and $x_i^2=R^2$ which is a sphere of radius $R$. After the resolution, the ALE 
space becomes a smooth Eguchi--Hanson space. This introduces an anomalous D--term in the D3 brane world--volume theory as in eq. (1).

The above setup reproduces exactly the theory in which there are cosmic D--term strings on the D3 brane world--volume. However, we have not yet identified the D--term string. In fact first 
we need to identify the D3 brane. There are two possibilities: it is either a D3 brane of type IIB string theory or a D5 brane wrapped on the resolved singularity[\MAL,\BER,\ZAF]. 
As we will see
below, the D3 brane is the latter. The D--term string can be an F or D string of IIB string theory or a D3 brane wrapped on the resolved singularity. In ref. [\POL], it was argued that
F1 and D1 branes cannot be stable if there are $M^4$--filling branes. This is because the end points of F1 and D1 branes on the D3 brane are charges and monopoles respectively and can be pair 
produced. As a result, F1 and D1 branes are not stable; they dissolve in the D3 brane and become electric and magnetic fluxes respectively. Moreover, the tensions of F1 and D1 branes 
are not proportional to the anomalous D--term or the blow--up
radius. Therefore, we cannot identify the D--term string with the F or D strings of type IIB string theory.  
We are left with the wrapped D3 brane. Note that its tension must be proportional to the blow--up radius since it wraps the resolved singularity.

We will take the D3 brane to be a D5 brane wrapped on the $P^1$ with radius $R$. Its tension is[\ZAF]
$$T_{D3}=T_{D5} \int_{S^2} \sqrt{det(G+B)} \eqno(15)$$
where $G$ and $B$ are the $P^1$ metric and NS--NS $B$ field on the $P^1$ respectively. We take $B=0$ for simplicity and find
$$T_{D3}={1 \over {(2 \pi)^6 g_s \ell_s^6}} 4 \pi R^2 \eqno(16)$$
This tension must be equal to the energy density on the brane world--volume that arises from the resolution of the singularity, i.e. the anomalous D--term. Thus we have
$T_{D3}=g^2 \xi^2$ where $g$ is the $U(1)$ coupling on the wrapped D5 brane given by
$${1 \over g^2}= {1 \over {(2 \pi)^3 g_s \ell_s^2}} \int_{S^2} \sqrt {det(G+B)}={R^2 \over {2 \pi^2 g_s \ell_s^2}} \eqno(17)$$
Then we find
$$\xi={R^2 \over {4 \sqrt 2 \pi^{7/2} g_s \ell_s^4}} \eqno(18)$$
The tension of the wrapped D3 brane is
$$T_{D1}=4 \pi R^2 T_{D3}={R^2 \over {4 \pi^3 g_s \ell_s^4}}= \sqrt{2 \pi} \xi \eqno(19)$$
Thus the wrapped D3 brane has exactly (up to a factor of $\sqrt {2 \pi}$) the tension of the D--term string. In addition it carries one unit of gauge flux just like a vortex due to the
world--volume Wess--Zumino coupling of the form $\int_{M^4} C_2 \wedge F_2$. (Here $C_2$ is given by the five--form $C_5$ that couples to a D3 brane reduced on the $P^1$.)
The wrapped D3 brane is a source of the Ramond--Ramond potential $C_2$ and therefore acts as a source of the gauge field $F_2$.
The above results lead us to identify wrapped D3 branes with D--term strings. Since, as we saw, the world--volume 
theory is exactly that of ref. [\COS], a solution that describes a string--like object with tension $\xi$ is necessarily given by eqs. (2)-(5).
This solution for the D--term 
string is the low--energy effective description of the wrapped D3 brane inside a wrapped D5 brane. Conversely, the wrapped D3 brane gives us the realization of the D--term string in 
IIB string theory. We also see the the charge (vortex number) of the D--term string is given by the number of the D3 branes that wrap the $P^1$. 
The stability of the wrapped D3 brane seems to follow from the wrapping which is a topological property. Unlike F and D strings, the D--term string cannot dissolve inside the D3 brane
because if it did we would be left with a two dimensional brane wrapped on the resolved $P^1$. Note that the brane configuration we have is a D3 brane inside a D5 brane. In flat space
this is T--dual to a D--string inside a D3 brane which is not stable. However, since both branes are wrapped on the $P^1$ this argument does not apply in our case.

Above we assumed for simplicity that $B=0$ on the resolved $P^1$. All our results generalize to the case with $B \not =0$ since from eqs. (17) and (19) we see that neither $G$ nor $B$ but 
only the sum $G+B$ is relevant for our results. For general $G$ and $B$ we find that $T_{D1}=\sqrt {2 \pi}\xi$ with
$$\xi= {1 \over {(2 \pi)^{3/2} \ell_s^2}}{1 \over g^2} \eqno(20)$$
where $g$ is given by eq. (17). We see that the D--term string tension is proportional to $1/g^2$ as it should be for a solitonic object. At the field theory level this is not
obvious since the gauge coupling $g$ and the anomaolous D--term $\xi$ are two independent parameters. However, we found that on the wrapped brane world--volume theory $\xi \sim 1/g^2$.

What happens when $R=0$ and $B \not =0$? If we do not resolve the singularity, we can still obtain a fractional D1 brane with nonzero
tension as long as $B \not=0$ by wrapping a D3 brane on the vanishing cycle[\BER,\ZAF,\JAU]. This fractional D1 brane seems to be stable (at least as stable as the D--term string) and describes 
a cosmic string. Taking the B--flux through the vanishing cycle to be $1/2$, i.e. $\int_{S^2} \sqrt B=(2 \pi \ell_s)^2/2$, we find
$$T_{D1}=\xi={1 \over {8 \pi^2}}{1 \over {g_s \ell_s^2}} \eqno(21)$$
This is not a D--term string since there is no D--term in the world--volume theory anymore. However, it must be a stable cosmic string described by a fractional D1 brane inside a
fractional D3 brane (which is a D5 brane wrapped on the vanishing cycle). We will call these fractional cosmic strings.
In ref. [\POL] it was argued that there are stable cosmic strings only when $\xi \not=0$ (or $R \not=0$ from our point of view). This is not surprising however, 
since we do not expect supergravity and field theory
to capture the physics of fractional branes which is purely string theoretical.

\bigskip

\centerline{\bf 4. Conclusions and Discussion}

\medskip

In this letter, we described the recently found D--term string solution as a D3 brane wrapped on a resolved conifold singularity. We considered the world--volume theory of a D3 brane
(which itself is a wrapped D5 brane) transverse to a conifold which is resolved by blowing up a $P^1$. We showed that the tension of this wrapped D3 brane is exactly that of the
D--term string. The stability of the cosmic string, i.e. the wrapped D3 brane seems to follow from its wrapping which is a topological property. We also argued that there must be a new 
type of cosmic string
which is described by a fractional D1 brane on the world--volume of a fractional D3 brane transverse to a conifold. This cannot be seen in the analysis of ref. [\COS] since its 
origin is purely string theoretical.

The cosmology of cosmic strings, especially that related to D--brane inflation has generated a lot of interest recently[\TYE]. These cosmic strings are remnants of $D3- \bar D3$ 
brane annihilations
which occur at the end of inflation and arise from the tachyon that appears at small interbrane distances. They are essentially D1 branes which are formed after tachyon condensation. 
D--term strings, on the other hand
arise from resolved singularities which are related to a new type of D--brane inflation[\DSS-\LAS] on fractional branes. In this type of brane inflation, the origin of inflation is 
the resolution
of an orbifold singularity[\BLO,\NEW]. This means that we should expect D--term cosmic strings also in these inflation models. It would be interesting to explore the cosmology of 
D--term strings in the framework of these new type of D--brane inflation models. 

One can also consider D3 branes at more complicated singularities. For example, if the D3 brane is transverse to a space that is locally a $Z_n$ ALE space fibered over a $P^1$, the 
world--volume theory is a quiver theory[\QUI] with gauge group $U(1)^n$. Then we can have $n-1$ anomalous D--terms (neglecting the overall diagonal $U(1)$) which lead to $n-1$ different 
D--term strings. These can be described by D3 branes wrapped on the $n-1$ $P^1$'s that are needed to resolve an $A_{n-1}$ type singularity. Each of these wrappings will give a different
stable cosmic string. Since the different $P^1$ can be of any radius
(corresponding to different $\xi_i$ for the anomalous D--terms) each of these cosmic strings will, in general, have a different tension. In fact, some of the cosmic strings may be
wrapped branes whereas others are fractional branes (with different fluxes over the $P^1$'s). Thus it is easy to find theories with a large number of stable cosmic strings
with different tensions.



\vfill

\refout

\end
\bye